# Chapter 6

# Cold Powering

A. Ballarino[1*], J.P. Burnet[1], D. Ramos[1], U. Wagner[1], S. Weisz[1] and Y. Yang[2]

[1]CERN, Accelerator & Technology Sector, Geneva, Switzerland
[2]University of Southampton, Southampton, UK

## 6    Cold powering

### 6.1    Overview

The electrical feed for the approximately 1700 superconducting (SC) circuits of the Large Hadron Collider (LHC) requires the transmission of more than ±1.5 MA of current from the power converters to the magnets. This is done via conventional copper cables for the room temperature path between power converters and current leads, and high temperature superconducting (HTS) and resistive currents leads for transfer to the 4.5 K liquid helium bath. Nb-Ti busbars operating in liquid helium at 4.5 K or in superfluid helium at 1.9 K provide the connection to the SC magnets. In the current LHC configuration, the power converters and the current leads are both located in underground areas, the former mainly in alcoves situated adjacent to the machine tunnel, and the latter in cryostats that are near the LHC interaction points and in line with the SC magnets. The 60 A power converters for the dipole orbit correctors are located in the tunnel, underneath the main dipole magnets. From each of the eight interaction points, power converters and current leads feed the magnets, occupying half of the two adjacent machine sectors. Some equipment in the tunnel is exposed to significant levels of radiation.

For the HL-LHC upgrade, novel superconducting lines (hereafter called 'links') are being developed to supply current to the magnets from remote locations [1]. The new electrical layout envisages the location of the power converters and current leads either in surface buildings or in underground areas, some hundreds of metres away from the tunnel. The transmission of the current to the magnets is performed via SC links containing tens of cables feeding different circuits. Each link would transfer in total up to about 150 kA. There are several benefits of remote powering via F and these can be summarized as follows.

- Access of personnel for maintenance, tests, and interventions on power converters, current leads, and associated equipment in radiation-free areas, in accordance with the principle of radiation protection that optimizes doses to personnel exposed to radiation by keeping them as low as reasonably achievable (ALARA).

- Removal of the current leads and associated cryostats from the accelerator ring, thus making space available for other accelerator components. In the baseline hardware layout of the HL-LHC interaction regions (IR) around P1 and P5, no space is available for current leads and cryostats in line with the magnets. Also, there is no space in the existing underground alcoves for locating the power converters feeding the HL-LHC circuits.

- Location of the power converters in radiation-free areas.

The ongoing programme focuses on the development of links to be integrated at LHC points P1, P5, and P7.

---

[*] Corresponding author: amalia.ballarino@cern.ch

### 6.1.1 Cold Powering Systems at LHC P1, P5, and P7

The Cold Powering System consists of the following.

- Power converters.
- Current leads, located as near as possible to the power converters in a radiation-free zone some hundreds of metres distance from the LHC tunnel. The leads are connected to the power converters via room temperature conventional cables.
- A dedicated cryostat (DFH), where the cold terminations of the leads are electrically connected to the cables in the link.
- A link, made of superconducting cables housed in a semi-flexible cryostat (DSH). The link electrically connects the leads to the magnet busbar.
- A cryostat (DF) in line with the magnets at the location where the link terminates in the LHC tunnel. In the DF cryostat, each cable of the link is connected to the Nb-Ti busbar feeding a magnet circuit. The helium cryogen required for the cooling of the Cold Powering System is supplied from this cryostat.
- Cryogenic instrumentation required for control, monitoring, and interlock functions as well as the electrical instrumentation needed for protection of superconducting components and current leads.

The Cold Powering System relies on cooling with helium gas, and the superconducting part of the system spans the temperature range from 4.2 K up to 35–50 K. The use of $MgB_2$ and HTS materials enables safe operation of the superconducting components, for which a temperature margin, intended as the increase of temperature that generates a resistive transition, of at least 10 K is guaranteed.

In the tunnel, vapour generated in the DF cryostat from a two-phase helium bath is conveyed inside the link cold mass [2]. The gas cools the SC cables in the link and warms up to about 17 K while absorbing the static heat load of the cryostat. The DSH cryostat includes a thermal shield actively cooled by forced flow of He gas taken from the tunnel at about 20 K. For the Cold Powering Systems at LHC P1 and P5, the two helium flows are mixed at the level of the DFH to produce the flow of He gas, at about 30 K, required for the cooling of the current leads. The system at LHC P7 does not need mixing – the helium flow in the superconducting link cold mass is sufficient for the cooling of the leads. The design of the current leads is such that the gas is recovered at room temperature at their warm end.

The development of the Cold Powering System for LHC P7 [1] is well advanced to date.

- Concepts [3] and prototype [4] SC cables have been developed and tested.
- Dedicated cabling machines conceived for production of long unit lengths of novel HTS or $MgB_2$ cables have been designed, assembled, and operated [5].
- A prototype Cold Powering System has been developed and tested [6].
- Integration studies in the LHC have been performed [7].
- New concepts of current leads and DFH cryostats optimized for easy transport and integration in the LHC underground areas have been elaborated. Detailed drawing activity for production of prototypes is being launched.

The design of the Cold Powering Systems minimizes the work done in the tunnel during installation of components. It also takes into account boundary conditions imposed by transport in the LHC underground areas. In particular, volumes of individual components have been minimized according to transport

---

[1] The SC link in P7 has recently been removed from the HL-LHC baseline; see LMC199 of 17 December 2014. However, the PDR refers to the baseline of October–November 2104, therefore we prefer to keep the description of the P7 SC link in this text. In addition this assures consistency with the first version of the PDR delivered to the FP7 office on 30 November 2014, and also avoids a revision of other parts of the PDR referring to the November 2014 baseline,

requirements, and the main components (link, i.e. DSH cryostat with the SC cables inside, current leads, and the DFH and DF cryostats) are integrated as complete and pre-tested assemblies. Activities after integration are limited to splicing between current leads and link and closure of the DFH cryostat; splicing between link and Nb-Ti busbar and closure of the DF cryostat; and connection of the conventional room temperature cables to current leads and power converters.

### 6.1.2 Superconducting link

The superconducting link is a semi-flexible transfer line, of a total length of up to about 500 m, which houses the SC cables connecting the cold end of the current leads to the Nb-Ti busbar of the magnet [1]. The transfer line consists of four corrugated concentric pipes that define the link cold mass, the actively cooled thermal shield, and the external vacuum insulation wall.

The number of SC cables contained inside the link and their operating current vary for the different Cold Powering Systems that are currently under study. Inside the cold mass of each link there are tens of cables rated at different DC currents ranging from a minimum of 120 A up to a maximum of 20 kA. The cables are grouped in the form of compact cable assemblies with a total current capability of up to about 150 kA.

At LHC P1 and P5 six superconducting links, three right and three left of each interaction point, are being considered for integration in the LHC machine:

- two links for the powering of the HL-LHC insertions (low-$\beta$ quadrupoles, D1 and corrector magnets);
- two links for the powering of the HL-LHC matching sections;
- two links for the powering of the LHC magnets in the arc[2].

The associated Cold Powering Systems replace the LHC cryogenic feedboxes (DFBX, DFBL, and DFBA). New current leads and cryostats (DFHX, DFHM, and DFHA), connected to the link in a radiation-free area away from the tunnel, are being developed.

Work until now has been focused on the systems for the HL-LHC upgrade, but the development is directly applicable to the Cold Powering Systems that are intended to replace the DFBA. The present baseline electrical layout envisages the installation of the power converters and current leads in surface buildings. This calls for the development of superconducting links, about 300 m long, including a vertical section of about 80 m.

At LHC P7 it is proposed to move power converters and current leads to an underground radiation-free gallery, which serves as access to the LHC ring (TZ76 gallery). Two superconducting links, each about 500 m long, are needed to connect the DF cryostat in the tunnel to the DFH in the TZ76 (see Section 6.1). The SC link will contain all cables feeding the LHC 600 A circuits that are currently powered via the DFBA and DFBM at P7. The cable assembly in the SC link cold mass requires 48 cables rated at 600 A, for a total current-carrying capability of about 29 kA.

The need of the Cold Powering Systems for the LHC circuits that are not part of the HL-LHC upgrade, i.e. those feeding either the magnets in the arc at LHC P1 and P5 or the magnets at P7, is driven by the Radiation To Electronics (R2E) requirements of the power converters. Integration with the LHC machine will be decided at a later stage of the project.

The present baseline proposal envisages using $MgB_2$ conductor in the longest part of the Links (from 4.2 K to 20 K), and HTS material (YBCO or Bi-2223) in the temperature range 20 K to 35–50 K. The potential low cost of the $MgB_2$ conductor and the possibility of cooling the superconducting link cold mass with He gas enable the development of Cold Powering Systems with improved performance. There is a higher temperature margin, with the benefit of safer operation and lower total exergetic cost of the refrigeration, with no extra cost with respect to the conventional Nb-Ti solution. Figure 6-2 shows a cross-section through the superconducting cable assembly proposed for the links powering the HL-LHC insertions at LHC P1 and P5. The total current transferred by these 44 cables is 165 kA. Six 20 kA cables are required for powering the low-$\beta$ insertion $Nb_3Sn$ quadrupole magnets and the Nb-Ti separation dipole, while the other cables feed corrector and trim circuits. Details of the cable assemblies developed for other Cold Powering Systems are presented elsewhere [1].

---

[2] SC links for the LHC magnet in the arc (change of DFBA) has been removed from the HL-LHC baseline (see LMC199 of 17 December 2104). However, for reason of coherence with the November 2014 baseline that is the reference for the present PDR (see previous footnote), the text has not been modified. The change will be recorded in the next version of the Technical Design Report due at the end of 2015.

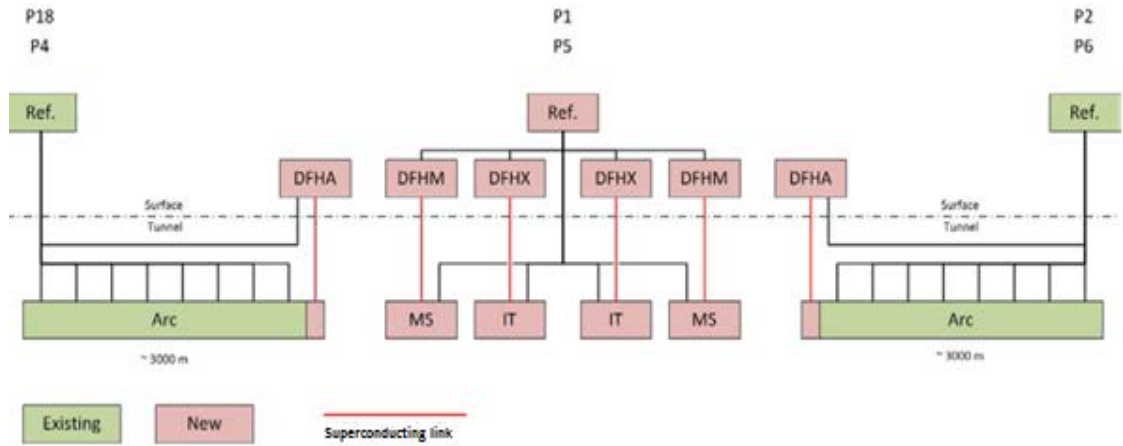

Figure 6-1: Superconducting links at LHC P1 and P5

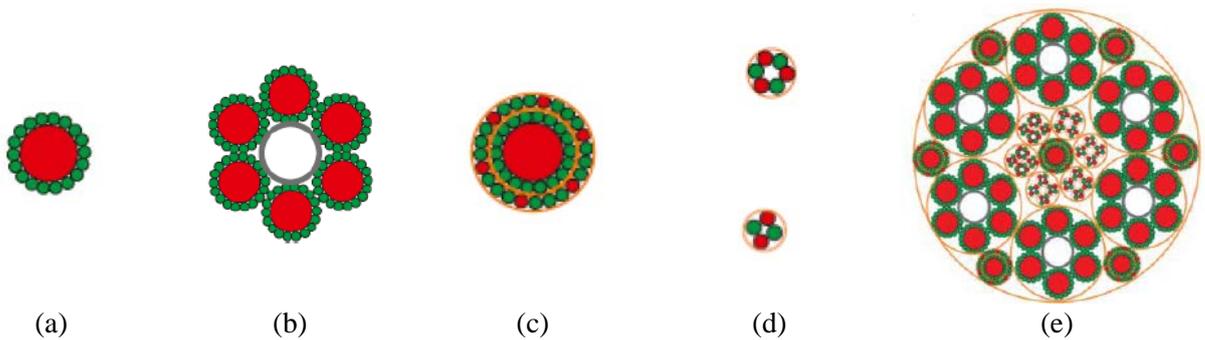

(a)      (b)      (c)      (d)      (e)

Figure 6-2: Cable assemblies for superconducting links at LHC P1 and P5. (a) Sub-unit of 20 kA cable, $\Phi$ ~6.5 mm; (b) 20 kA cable, $\Phi$ ~19.5 mm; (c) concentric $2 \times 3$ kA cable, $\Phi$ ~8.5 mm; (d) 0.4 kA cable (top) and 0.12 kA cable (bottom), $\Phi$ <3 mm; (e) 165 kA cable assembly for LHC P1 and P5 ($6 \times 20$ kA, $7 \times 2 \times 3$ kA, $4 \times 0.4$ kA, $18 \times 0.12$ kA), $\Phi$ ~65 mm. The cables are made of copper stabilizer (red) and $MgB_2$ wire (green).

The cable (twisted-pair superconducting tapes or wires) and cable assemblies developed for LHC P7 are optimized for the transport of current in the 1 kA range. Details of the cable concepts and results from tests performed under nominal operating conditions are presented elsewhere [4]. The cable assemblies are incorporated in a semi-flexible cryostat of the Cryoflex® type. The present baseline, which is to be confirmed through ongoing integration studies and tests, envisages integration in the LHC tunnel of the cryostat with the cable assemblies already pulled in at the surface. To limit the risks associated with high-current resistive joints operated in a helium gas environment, the cables are planned to be assembled in one single unit length with no splices between cables inside the link.

The main achievements to date are:

- the development of the first $MgB_2$ powder-in-tube (PIT) round wire with electrical and mechanical performance that permits its use in high-current cables – work done in collaboration between CERN and Columbus Superconductors, Genova;

- the test of a 3 m long superconducting link of the type needed at LHC P7 [6];
- the design of a Cold Powering System optimized for integration at LHC P7 [6];
- the successful development and test of the first $2 \times 20$ m long cable made from $MgB_2$ round wire operated successfully up to 20 kA at 24 K [8].

Figure 6-3 shows the test station designed and successfully operated at CERN for the test of superconducting links up to 20 m long. As in the final configuration, the cables are cooled by the forced flow of helium gas operating at any temperature from about 5 K to 35 K. Temperatures of up to 70 K can be achieved, enabling appraisal of cables made from different types of conductor.

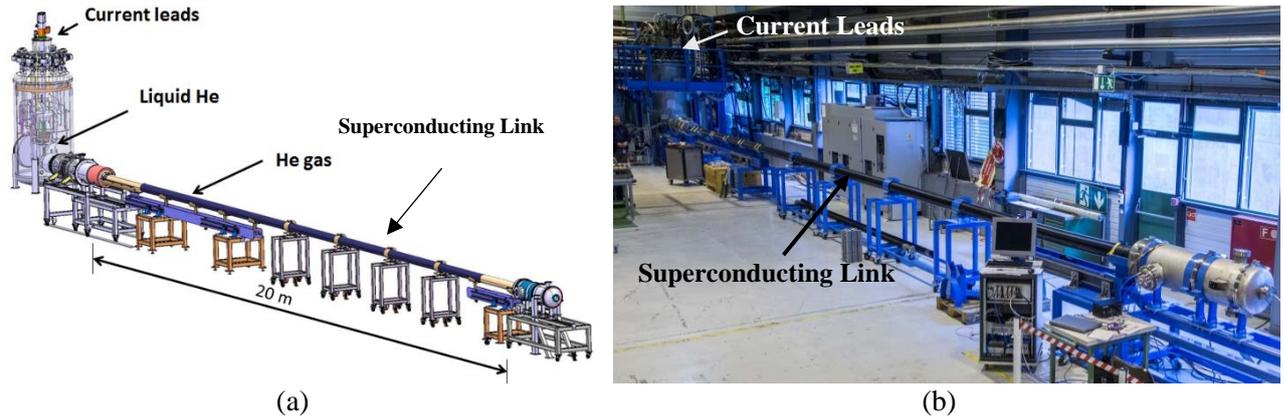

Figure 6-3: Two views of the test station designed and operated at CERN for the test of superconducting links up to 20 m long. The cables are cooled by forced flow of helium gas operating at any temperature from about 5 K to 35 K. Temperatures of up to 70 K can be achieved, enabling appraisal of cables made from different types of conductor.

## 6.2 Cold Powering System design

In essence the Cold Powering System for the powering of the LHC magnets by superconducting links is a semi-flexible cryostat extending over a few hundred metres. The novel design has to face several challenges never encountered previously.

6.2.1 Cryostat for the superconducting link

The superconducting link cryostat (DSH) must maintain a stable and well-defined cryogenic environment within which the superconducting cables are cooled by the forced flow of helium gas. Unlike the liquid helium-cooled superconducting busbars in the LHC machine, the cryogenic stability of the superconducting cables depends more critically on the cooling efficiency in a helium gas environment within the superconducting link cold mass. The basic cryostat structure consists of an inner vessel surrounded by an actively cooled thermal shield and enclosed by an outer vessel as room temperature vacuum envelope. At present the reference design of the cryostat is analogous to Nexans' four-tube coaxial Cryoflex® transfer line, which can be manufactured and delivered in one continuous length compatible with the cold powering requirements. It is essential to minimize the number of splices between cables inside the SC link cold mass due to their multiple circuit complexity and cooling challenges for high-current resistive joints cooled by gaseous helium. While cryostat flexibility is a crucial design aspect, it is recognized that the mechanical support to the vertical section at LHC P1 and P5 requires additional consideration. As part of the HL-LHC design studies, pre-prototype cryostats in 5 m and 20 m lengths have been procured from Nexans and tested at CERN [8] and at the University of Southampton [6]. The working experiences have been positive so far and a 60 m long cryostat will be procured in the coming months for integration tests in configurations similar to those in the LHC underground areas and for horizontal and vertical evaluation.

The gaseous helium in the cold mass of the link provides cooling for (i) steady-state heat load of the radial heat in-leak (conduction and radiation) from the surrounding thermal shield and heat conduction along the link and the inner vessel wall from the warmer end at 20 K; and (ii) transient heat load due to local disturbances in the link and/or thermal/vacuum instability in the cryostat. Under nominal operating conditions, the radial heat in-leak is dominant, estimated conservatively at below 0.2 W/m for a thermal shield temperature of ~60 K without multilayer insulation. The thermal shield will be actively cooled by flowing helium gas with an inlet temperature at 20 K. Further discussions on the cooling requirement can be found in Ref. [2].

### 6.3 Interfaces to the superconducting link

The link has a colder (4.2 K) interface to the LHC machine at one end and a warmer (20 K) interface to the current leads at the level of the DFH cryostat. The two interfaces are an integral part of the Cold Powering System due to electrical continuity and synergy in the cooling arrangements.

The present LHC current leads consist of a self-cooled HTS section and a 20 K helium gas-cooled resistive copper section [9]. The former is connected to the Nb-Ti busbar in helium and the latter extends from the warm end of the HTS at 50 K to room temperature. The Cold Powering System design seeks the integration of the different helium gas flows at the interface of the superconducting link and current leads.

#### 6.3.1 Electrical interface between the superconducting link and the current leads

Each of the multiple cables inside the superconducting link is spliced and connected to its corresponding current lead inside the DFH interconnection cryostat. The cryostat must allow easy access for making the electrical connections after the deployment of the superconducting link cable assembly in the semi-flexible DSH cryostat. The primary design focus is the reliable and secure handling of the SC cables via robust tooling and procedures. In addition, the design ensures in situ completion of low resistance joints between the SC cables in the link and the current leads. Effective cooling of the resistive joints via the helium gas inside the link also requires integrated heat transfer features.

There are additional considerations to address specific circumstances for the systems at P1, P5, and P7. For example, the SC link at P7 terminates in the TZ76 underground gallery. The present configuration of TZ76 imposes stringent space restrictions for component transportation and installation. These constraints have been satisfied in the present baseline design where a stand-alone interconnection cryostat is connected to twelve discrete units of four × 600 A current lead assemblies via HTS cables housed in twelve slim transfer-line type cryostats. A prototype of the Cold Powering System at LHC P7 will be manufactured and tested in the SM18 in 2015. The existing test station incorporating the 20 m Cryoflex® line (see Figure 6-3) will be used for this purpose.

#### 6.3.2 Cryogenic interface between the superconducting link and the current leads

The interface between the superconducting link and the current leads involves several cryogenic aspects. First of all, the continuation of the helium gas from the superconducting link into the current leads and its mixing with additional cooling gas if necessary will be assured. The present cooling proposal for the systems at P1 and P5 [2] uses the warm exit of the helium gas cooling the SC link thermal shield as the supplementary coolant for the current leads. Since the exit temperature of the shield cooling gas always exceeds the warm boundary condition of 20 K for the SC cable in the link, this option implicitly imposes hydraulic separation of the helium spaces of the link and the leads. This is likely to be the scenario for P1 and P5, where the superconducting link cold mass requires ~1 g/s while the current leads require ~10 g/s for a total transfer of about 150 kA.

The cooling requirements for the SC link and current leads at P7 are well matched, with 1.5 g/s for the DSH cold mass and 1.4 g/s for the DFH. Therefore, there is a compelling case for using only the helium gas from the SC link cold mass for cooling the current leads. The elimination of hydraulic separation/re-mixing is an attractive proposition for aspects of both assembly and installation.

It is envisaged that the superconducting link cryostat (DSH) and current leads share a common vacuum space. A Paschen scenario will be avoided by ensuring that all of the high-tension side is surrounded by helium cooling gas at about 1 bar.

### 6.3.3 Control

The stable operation of the Cold Powering System relies on maintaining two temperature boundary conditions, i.e. 20 K at the splice terminations between the superconducting link and the current leads and about 35 K to 50 K at the cold end of the resistive section of the current leads. The former is controlled by a heater to generate the required helium boil-off from the DF cryostat in the LHC tunnel while the latter is controlled by a valve at the helium warm exit of each current lead. If a single helium flow is adopted for P7, then an appropriate override should be devised for the two controllers to work correctly in tandem. Specifically, if more flow is required by the current leads, the boil-off heater must allow the temperature in the DF to drift below the set value of 20 K. Conversely, if higher boil-off is necessary for the cooling of the interconnections, a pressure-controlled cold bleed will be used to discharge the excess in order to avoid over-cooling of the current leads.

## 6.4 Interface to the LHC machine

### 6.4.1 General

The DF cryostat interfacing the superconducting link to the magnets cold mass performs the role of electrical, cryogenic, and mechanical interface. It includes:

- the required connections to the LHC cryogenic distribution line (QRL);
- a saturated liquid helium bath for the electrical splices between the cables in the link and the Nb-Ti cables;
- a hydraulic separation with respect to the superfluid helium bath cooling the magnets;
- the instrumentation required for cryogenic process control.

Different variants of DF cryostats are necessary for the Cold Powering Systems under study. The DF cryostats are vacuum-insulated and equipped with an actively cooled thermal shield wrapped in multilayer insulation. Vacuum barriers are foreseen to separate the insulation vacuum of the DF cryostat from that of the link, in order to allow interventions on either piece of equipment without the need for vacuum conditioning of the full system.

### 6.4.2 Interface cryostat for the HL-LHC insertions

A continuous cryostat of approximately 60 m in length is foreseen to house the magnets from Q1 to D1 in a common insulation vacuum, with the interface cryostat to the link (DFX) located at its non-IP extremity, after D1. This being the most suitable location from the machine optics point of view, it implies on the other hand that the DFX vacuum vessel and respective supports must be designed to withstand an axial force of up to 10 tonnes, which is induced by unbalanced atmospheric pressure. The DFX will include a jumper to the QRL with helium piping for the supply of both the superconducting link and part of the continuous cryostat.

The DFX may either be designed as an independent cryostat or as a service module integrated in the D1 cryostat. The choice between these two configurations will be made not only on the basis of integration and engineering considerations, but most importantly taking into account the need for minimizing the residual radiation doses to personnel during specific interventions that may occur in the lifetime of the HL-LHC machine, such as the exchange of a magnet.

### 6.4.3 Interface cryostat for the matching sections

The most compact solution for the Cold powering System of the matching sections is to include the connection to the link in the service module of the magnet cryostat (QQS). From there, a first link cooled by supercritical helium is routed up to the DFM cryostat that replaces the present LHC DFL feedbox. All cables are then gathered inside the DFM into a main link connecting the DFM to the current leads at the surface. The supercritical helium arriving from the link in the tunnel is expanded inside the DFM to generate liquid for the splices and gas for the cooling of the SC link joining the surface. As such, this concept does not require a connection to the QRL at the level of the DFM.

6.4.4    Interface cryostat for the arcs

The DFBA LHC feedboxes comprise the so-called current modules, housing the current leads, and a shuffling module that serves as an interface to the Q7; all being supported in a stiff beam which spreads the axial force from atmospheric pressure to the tunnel floor. The most cost effective approach consists in exchanging the existing current module with a new DFA cryostat, profiting from the existing interfaces to the arc cryostat and the QRL. In this way, no interventions are required on the shuffling module and on the support beam. The DFA includes the connection to the QRL and provides the helium supply for the link. The DFA also contains two 13 kA current leads, used for energy extraction of the main dipole circuit, which will remain in their current location. Because they require a precise control of the liquid helium level, these leads will be cooled by a separate helium bath. In the particular case of LHC P7, a supercritical helium cooled link connects the Q6 to the DFA.

**6.5    Integration of the Cold Powering Systems in the LHC machine**

The constraints for integration of the Cold Powering Systems at LHC P7, P1, and P5 are very different, and each of these cases requires a specific analysis:

- P7 is below the border between France and Switzerland, near the Lycée International of Ferney Voltaire, and surface buildings cannot be envisaged in this environment;
- P1 is equipped with a large service shaft and the LHC tunnel is enlarged in the corresponding straight section;
- P5 underground conditions are very restrictive, with a service shaft that is small and off-centre, and with standard LHC tunnel around the collision region.

The following sections summarize the solutions presently retained for installing the Cold Powering Systems in these different areas.

6.5.1    Integration of the Cold Powering Systems at LHC P7

The insertion at P7 houses the betatron collimation system, where a large fraction of the beam losses is intercepted. This induces a radiation level that is a long-standing issue regarding single event upset in the electronics devices installed in the area. Additional shielding to protect the RR was installed in 2008 [10]. The uninterruptible powering system (UPS) was also relocated in 2008 [11]. All sensitive elements were relocated out of UJ76 during Long Shutdown 1 (LS1) [12]. The power converters located in the service caverns RR73 and RR77 are the next source of concern as both the energy and intensity of the circulating beam will increase in the future. The solutions envisaged require either to increase the radiation tolerance or to relocate these power converters, and the Cold Powering System provides a practical solution for the second option.

As mentioned above, the geographical situation of P7 strongly constrains the installation of surface buildings, and this is the reason why the access shaft, which dates from the LEP construction era, is located about 400 m from the machine tunnel. Figure 6-4 gives an overview of the corresponding underground premises. The long gallery TZ76, which was already used to relocate equipment originally installed in the UJ76 cavern, provides a radiation-free area to install the power converters after their removal from RR73 and RR77. The red and yellow lines in Figure 6-4 indicate the routing of the Cold Powering Systems associated with arc 6–7 and arc 7–8, respectively. The total length of each line is about 500 m, with 250 m along the beam line in the LHC tunnel. Section 6.5.2 presents the integration of the Cold Powering System along the different portions of these routings, starting from the RR toward the TZ service gallery.

The RR73/R77 alcoves contain the LHC DFBAM/N feedboxes that provide the warm to cold transition via the current leads. These DFBs will be replaced by the DFAM/N interfaces between the SC link and the Nb-Ti busbar of the magnets.

The links and the helium gas warm recovery lines can be installed close to the ceiling of the LHC tunnel in the long straight sections R74 and R771.

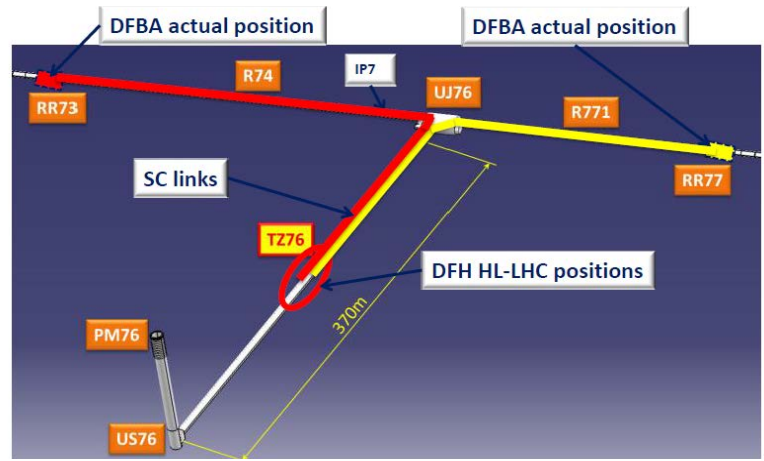

Figure 6-4: Underground layout at LHC P7

The passage of the links from the LHC tunnel to the TZ76 service gallery is the most delicate area regarding the installation of the Cold Powering Systems at LHC P7. Two alternatives are presently under consideration. The first involves a large number of rather sharp bends of the SC link to comply with the geometry of the UJ76 junction area (see Figure 6-5). The second relies on the drilling of two long ducts (each ~19 m long) to simplify the routing of the SC link (see Figure 6-6). A 60 m long cryostat should become available in spring 2015 to test the handling and installation of a SC link in order to decide between these two options.

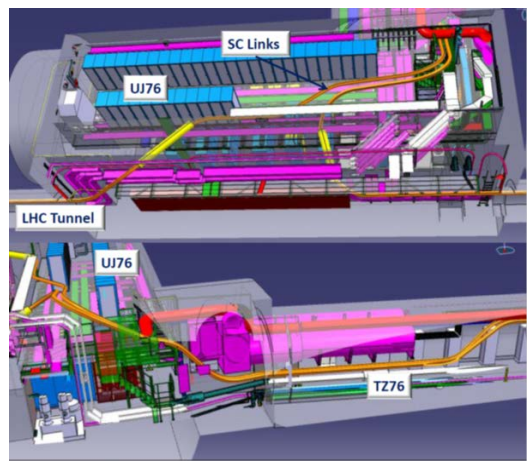

Figure 6-5: Passage through UJ76 with routing of the links in the UJ76 area

The two links and the two He return lines can be installed on the upper part of the TZ76 service gallery as shown on Figure 6-7(b). The routing in this configuration would be about 500 m until one reaches an equipment-free area where the current leads and the relocated power converters can be installed. Figure 6-7(a) gives a schematic view. The precise dimension and shape of the DFH cryostats are being defined. Space has been allocated for the power converters and their associated cooling units.

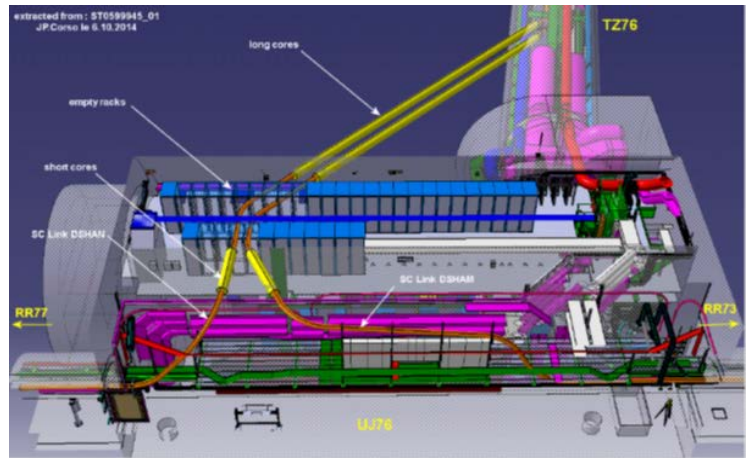

Figure 6-6: Passage of links through UJ76 with additional ducts

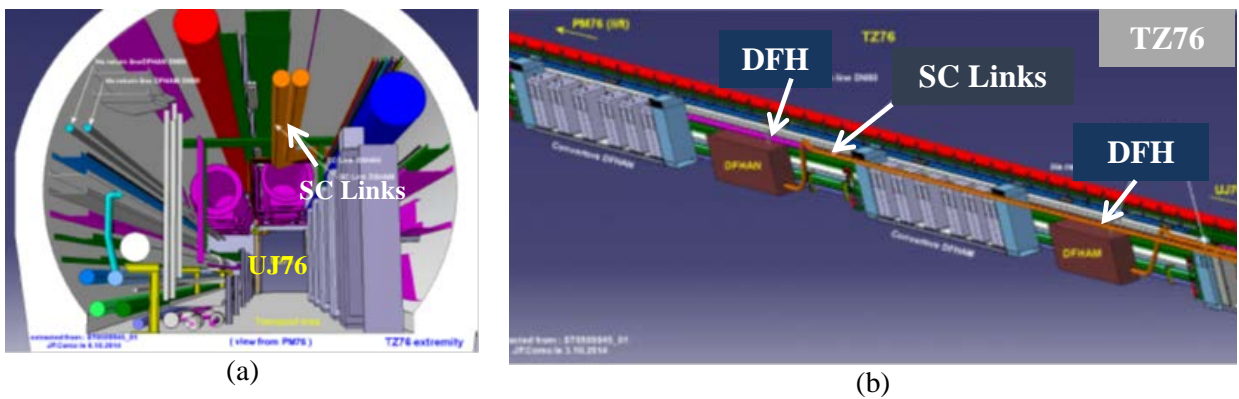

Figure 6-7: (a) Routing of the two links and the two He return lines in the service gallery. (b) Installation of the DFH and of the power converters in TZ76.

6.5.2   Integration of the Cold Powering Systems at LHC P1

P1 houses high luminosity insertions that are totally redesigned to meet the HL-LHC performance requirements. The inner triplet quadrupoles, the D1–D2 separation dipoles, and the matching sections will be replaced. Much higher current will be required to power the new magnets. The present underground premises at P1 do not provide enough space for the corresponding power converters; and the objective of the Cold Powering System is to allow the installation of these new power converters at the surface. There is a large service shaft (PM15) at P1 that can be used to route several SC links, and those powering the new low-$\beta$ triplets and D1 have a quite simple path as shown in Figure 6-8.

The D2 separation dipole and the stand-alone quadrupoles of the matching section are presently powered through a Nb-Ti superconducting link with distribution feedboxes (DFBLA and DFBLB) and power converters located in the alcoves located at about 250 m distance from the tunnel on both sides of IP1 (RR13 and RR17). The new Cold Powering Systems include DFM cryostats between the Nb-Ti magnet busbars and the $MgB_2$ cables that would be installed in place of the present DFBLA and DFBLB. The links will then be routed along the beam lines in the long straight sections until reaching the junction caverns (UJ13 and UJ17, left and right, respectively, of IP1) and then running to the surface as shown in Figure 6-8. The precise position of the links in the long straight section is not defined yet as they will need to pass across the crab cavities area that is still being designed. However, as the LHC tunnel is enlarged to a 4.4 m diameter around P1, no space difficulties are expected.

The RR13 and RR17 also house feedboxes (DFBAA and DFBAB) and power converters for arcs 8-1 and 1-2, respectively. Radiation levels in these areas will increase with luminosity and may become an issue. Moreover, the HL-LHC involves new equipment in the high luminosity insertions that, in turn, will require additional space. While the impact of the radiation level and the need for shielding are being assessed, there are clear incentives to relocate as many as possible of the present power converters from RR13/RR17 to the surface. If this is the case, the removal of the DFBAs would free space for the installation of new DFA cryostats, with the current leads for energy extraction of the main dipole circuit and the connections to the QRL remaining at their current locations. The links would then follow the same path to the surface as those discussed above for D2 and the matching section quadrupoles.

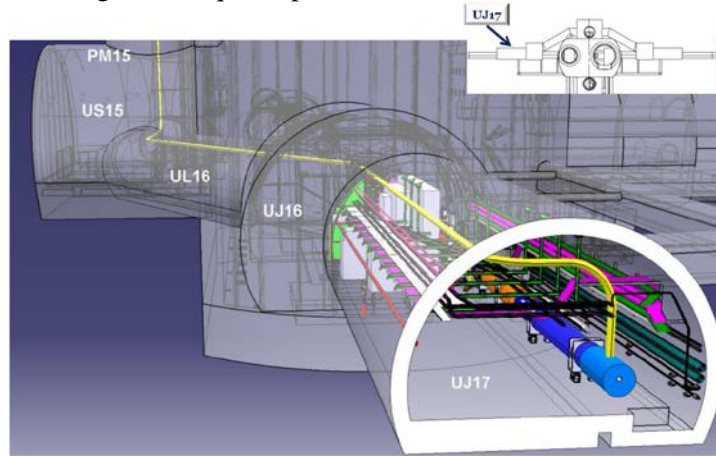

Figure 6-8: Routing of the links to surface for the low-$\beta$ triplet right of LHC P1

The detailed integration of the power converters in a surface building is not yet available, but it will involve an extension of the SD15 building at the top of PM15 with a total size that has to account for the total number of relocated power converters and for other additional HL-LHC related equipment.

6.5.3    Integration of the Cold Powering Systems at LHC P5

The elements of the Cold Powering Systems to be installed for HL-LHC at P5 are the same as those for P1. However, the environment at P5 is different. The tunnel diameter in the straight sections is 3.8 m, and nearly all of the space is already allocated. The service shaft (PM56), located to the right of P5, is a small shaft dedicated to personnel access only. Therefore, a totally different routing to the surface must be envisaged for the links compared with the solution retained for P1.

Vertical shafts would allow routing the links from the LHC machine to the surface. They can be done with a standard excavation method (auger drilling) providing a 0.4 m internal diameter duct that is sufficient for housing one SC link and its supporting system. As it is a top-down drilling technique, there is a tolerance of about 0.5 m on the position of the shaft when it reaches the LHC level some 90 m below ground. This would be problematic if the target is the narrow machine tunnel, but it is acceptable if the shaft has to reach enlarged areas as the RRs or UJs, which are 10.40 m and 8.40 m wide, respectively. The position of these enlarged areas at P5 is indicated in Figure 6-9. Vertical shafts above the UJ53 and the UJ57 would be allocated to the links for the new low-$\beta$ triplets and D1 separation dipoles. Those above the RR53 and the RR57 would be allocated to the links for the stand-alone magnets of the matching sections. If the power converters for the arcs are also relocated to the surface, shafts would be drilled above the RR53 and the RR57 for the corresponding links.

New surface buildings will be required to house the power converters. The optimization process between small local buildings or larger ones, connected through trenches to regroup the power converters, has yet to be done. A global optimization should take account of all new equipment required for the HL-LHC. The crab cavity system is facing a similar problem and it should be mentioned that a study is ongoing for new underground premises to cope with the space requirements of the cryogenics cold boxes, the crab cavities, and the powering equipment.

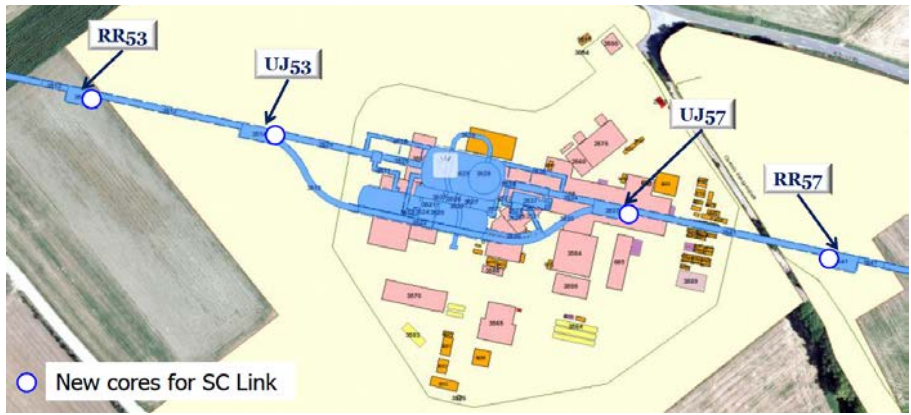

Figure 6-9: Layout of underground premises and of surface buildings at P5

## 6.6 Powering layout

Figure 6-10 shows the powering layout of the present LHC inner triplets. This powering scheme consists of a nested circuit with three power converters and free-wheeling protection circuits [13].

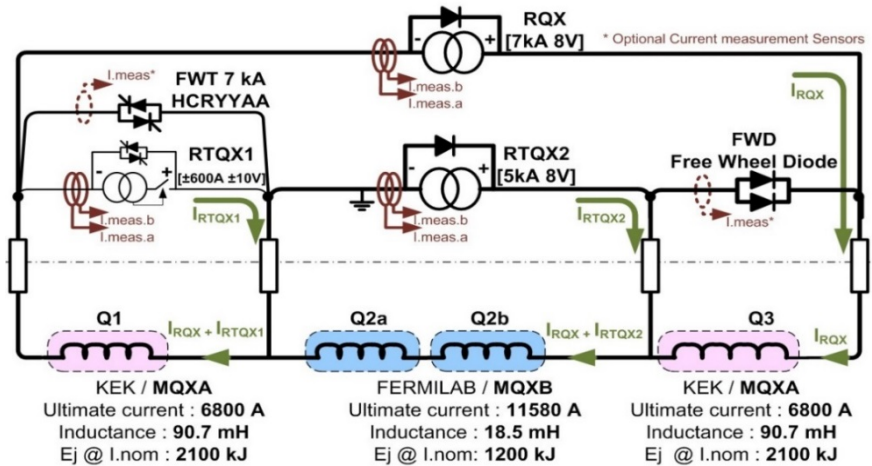

Figure 6-10: Powering layout of the present LHC inner triplets

It requires a dedicated control system to decouple the control of the power converters [14]. At the beginning of the LHC operation, this specificity generated longer downtimes with respect to the other electrical circuits. Thanks to the experience gained and the development of diagnostic tools, the system has currently reached a good level of reliability. It should be noted that the ramp-down time of these circuits, which defines the minimum time before a beam injection, had to be slowed down with respect to that obtained with the free-wheeling process. This was done in order to avoid trips of the power converters generated by the nested configuration.

The baseline powering layout of the HL-LHC inner triplets is shown in Figure 6-11. The low-$\beta$ quadrupoles (Q1, Q2, and Q3) are powered via two main circuits, each equipped with one trim power converter. The two Q2 units are powered in series with a 200 A trim converter on Q2b. Q1 and Q3 are powered in series with a 2 kA trim converter on Q3. The separation dipole (D1) and the corrector magnets are individually powered. This layout: i) provides maximum flexibility for beam optics; ii) simplifies the powering scheme with respect to the layout in Figure 6-10 with the advantage of a reduction of the mean time to repair (MTTR) for interventions on the power converters. It should be noted that the total current transferred for feeding each triplet increases from about 40 kA in the present LHC scheme to above 150 kA for the HL-LHC configuration.

The possibility of powering all low-$\beta$ quadrupoles in series, with dedicated trims, was considered. However, the layout in Figure 6-11 was preferred because of the smaller inductance per circuit and therefore reduced challenges for magnet quench protection. The powering of each low-$\beta$ quadrupole magnet via a separate circuit was also not retained because of the related higher cost of the powering equipment – power converters and Cold Powering System.

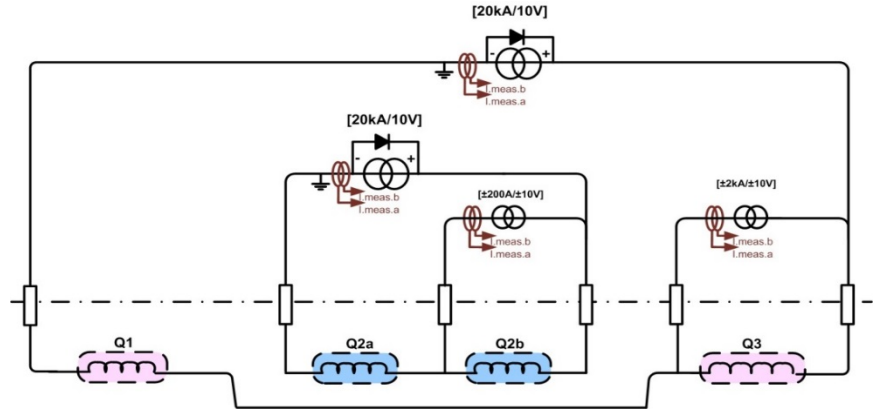

Figure 6-11: Powering layout of the HL-LHC inner triplets

The powering layout of the HL-LHC matching sections has also been reviewed. In the present LHC matching sections, the quadrupole magnets for Beam 1 and Beam 2 have a common return current lead. This choice was made in order to reduce the number of current leads from four – for individually powered magnets – to three. This electrical coupling puts constraints on the current settings of both circuits with the consequence of reducing the flexibility for different beam optics configurations. For this reason, in the HL-LHC matching sections it is proposed that quadrupoles Q4, Q5, and Q6 are individually powered with two current leads per circuit.

In the present LHC matching sections one-quadrant power converters are used for all of the quadrupoles. The decrease of the magnet current is done with a free-wheeling process, and the ramp rate depends on the time constant of the circuit, which is in turn defined mainly by the inductance of the circuit. With the present configuration, the beam squeeze process takes up to 20 minutes – and for the HL-LHC, the squeeze process would be even longer due to the wider range of $\beta^*$. The use of two-quadrant power converters (with bipolar voltage) is being considered for Q5. This would allow a faster decrease of the quadrupole current and therefore a reduced time for the beam squeeze process.

## 6.7 Power converters

The LHC was built with modular power converters to facilitate maintenance and integrate the redundancy principle [15] – redundancy was included in all LHC power converters rated at currents above 600 A. This has proven to be a real asset during operation. The $n + 1$ redundancy allows the power converters to be run even with one module in fault. The advantages are the following: i) in the case of a fault, only one sub-converter is not operational and, in most cases, the fault does not generate a beam dump; ii) the LHC can run with some faulty sub-converters in the machine and all interventions for repair can be performed during a machine technical stop. With the exception of the dipole magnets, switch-mode technology was chosen for the LHC power converters in order to minimize their size and assure low output voltage ripple. All LHC power converters rated at currents above 120 A are water-cooled, with the advantage of a reduced size of the hardware. All these design principles will be maintained for the new power converters of the HL-LHC magnets.

### 6.7.1 Performance of the power converters

The HL-LHC power converters will be regulated in a closed current loop. Those for Q1, Q2, and Q3 will be of the highest precision class, like the LHC main dipole circuits. The stability of the low-$\beta$ power converters will be at the level of $1 \times 10^6$ ($10^{-6}$ of nominal current) – with an uncertainty of $\pm 1$ ppm [16]. The stability of the other power converters will be of the order of $10 \times 10^6$. The achievement of this level of stability does not require any new developments. All hardware developed for the LHC power converters will be re-used, e.g. for the 20 kA DCCT and 22 bits ΣΔ ADC.

For the control of the power converters, the powerful function generator and controller (FGC) system, which allows control of all LHC power converters with the same hardware and software [17], is proposed. Its limitation is within the WORLDFIP bus, which has a speed that is limited to 2.5 Mbits/s. The replacement of this bus by an Ethernet type bus is being considered.

### 6.7.2    Location of the power converters

The present LHC power converters are installed in underground areas. Of the 1710 total units, 1065 are exposed to radiation. During machine operation up to 2013, the power converters generated a number of beam dumps due to single event effect (SEE). The faults due to SEE represented about 20% of the total power converter faults. A R2E programme was launched in 2010 to mitigate radiation issues for the whole LHC machine. In this framework, all power converters connected to the present DFBX were relocated to reduce their exposure to radiation. More shielding was added inside the RR alcoves to reduce particles fluences. A new radiation-tolerant version of the FGC system, called FGClite, was developed for integration in the machine in 2016. The 600 A 4 kA, 6 kA, and 8 kA power converters are being redesigned to be radiation-tolerant. After the full deployment of this programme – presently foreseen for 2018 – all power converters will able to operate in the RR alcoves and withstand the radiation levels foreseen for the HL-LHC.

According to the present baseline, the power converters for the powering of HL-LHC magnets will be installed in surface buildings. The ongoing development of radiation-tolerant power converters aims at operating the power converters for the circuits fed via the DFBA in the RR alcoves, where they are currently located.

### 6.7.3    Power converters for the HL-LHC

The list of power converters for HL-LHC is reported in Table 6-1. The list of magnets and corresponding power converters needed for HL-LHC is reported in Table 6-2. In the present LHC machine, 8kA, 600A, and 120A power converters are already in operation.

Table 6-1: List of new power converters for inner triplet and matching section magnets

| Power Converter | Current [kA] | Voltage [V] | Quadrant | Quantity IP side | Spare | Total quantity |
|---|---|---|---|---|---|---|
| Type 1 | 20 | <20 | 1 | 4 | 2 | 18 |
| Type 2 | 18 | <20 | 1 | 2 | 2 | 10 |
| Type 3 | 14 | <20 | 1 | 2 | 2 | 10 |
| Type 4 | 8 | <±10 | 2 | 4 | 2 | 18 |
| Type 5 | ±4 | <±10 | 4 | 15 | 6 | 66 |
| Type 6 | ±600 | <±10 | 4 | 1 | 2 | 6 |
| Type 8 | ±200 | <±10 | 4 | 9 | 6 | 42 |
| Type 9 | ±120 | <±10 | 4 | 8 | 6 | 38 |
| Total | | | | 45 | 28 | 208 |

The HL-LHC will require the development of two types of new power converters: i) the one-quadrant converter 20 kA/10 V, which will be based on an extension of the 8 kA LHC power converter family; ii) the four-quadrant converter ±4 kA/±10 V, which will be based on the present topology of 600 A LHC power converters.

The HL-LHC could require new topologies of two-quadrant power converters rated at 8 kA/±10 V. This development is mandatory if the squeeze time process needs to be reduced.

Also, by replacing the present 13 kA/18 V power converters of the main quadrupole circuits with two-quadrant power converters 13 kA/±18 V, the ramp down of the machine can be reduced by about 30 minutes. This upgrade would increase the global availability of the machine by about 4%.

Table 6-2: Magnets at each side of LHC P1 and P5

| Optics | Magnet | Operating current [kA] | Inductance [mH] | Power converters [kA] | Uncertainty ppm of max current with weekly calibration | ½ hour stability ppm of max current |
|---|---|---|---|---|---|---|
| MQXF Q1-Q3 | MQXF | 17.46 | 170 | 20 | ±1ppm | ±1ppm |
| Q3 | Trim Q3 | ±2 | 85 | ±3.2 | ±10ppm | ±5ppm |
| Q2a-Q2b | MQXF | 17.46 | 145 | 20 | ±1ppm | ±1ppm |
| Q2b | Trim Q2 | ±0.3 | 72.5 | ±0.8 | ±100ppm | ±10ppm |
| CP | MCBX | ±2.42 | 25 | ±3.2 | ±10ppm | ±5ppm |
|  | MQSXF | 0.182 | 1600 | ±0.2 | ±100ppm | ±10ppm |
|  | MCTSXF | 0.167 | 600 | ±0.2 | ±100ppm | ±10ppm |
|  | MCTSXF | 0.157 | 150 | ±0.2 | ±100ppm | ±10ppm |
|  | MCDXF | 0.139 | 300 | ±0.2 | ±100ppm | ±10ppm |
|  | MCDSXF | 0.139 | 300 | ±0.2 | ±100ppm | ±10ppm |
|  | MCOXF | 0.12 | 400 | ±0.2 | ±100ppm | ±10ppm |
|  | MCOSXF | 0.12 | 400 | ±0.2 | ±100ppm | ±10ppm |
|  | MCSXF | 0.132 | 180 | ±0.2 | ±100ppm | ±10ppm |
|  | MCSSXF | 0.132 | 180 | ±0.2 | ±100ppm | ±10ppm |
| D1 | MBXF | 11.8 | 36 | 14 | ±10ppm | ±5ppm |
| D2 | MBRD | 12 | 37 | 14 | ±10ppm | ±5ppm |
|  | MCBRD | ±3.2 | 12 | ±4 | ±10ppm | ±5ppm |
| Q4 | MQYY | 15.65 | 6 | 18 | ±10ppm | ±5ppm |
|  | MCBYY | ±3 | 12 | ±4 | ±10ppm | ±5ppm |
| Q5 | MQY | 4.21 | 74 | 8 | ±10ppm | ±5ppm |
|  | MCBY | 0.088 |  | ±0.12 | ±100ppm | ±10ppm |
| Q6 | MQML | 5.39 | 21 | 8 | ±10ppm | ±5ppm |
|  | MCBC | 0.1 |  | ±0.12 | ±100ppm | ±10ppm |

## 6.8 Radiation-tolerant power converters

The power converters currently in the RR alcoves will be replaced with radiation-tolerant converters. This development concerns the 600 A and the 4 kA, 6 kA, and 8 kA families. The replacement is planned to take place during Long Shutdown 2 (LS2). The new power converters will be able to withstand the doses and the fluences expected after the HL-LHC upgrade.

The present 60 A converters will not withstand the doses estimated during HL-LHC operation. They were designed for tolerating a maximum total dose of about 50 Gy, and the power converters placed in or close to the matching sections will receive a dose of up to 32 Gy/year. These converters will be replaced with new ones designed for withstanding a total dose of 200 Gy. This target corresponds to the maximum dose that can be tolerated by a design based on commercial off-the-shelf (COTS) components. A rotation between highly exposed and less exposed power converters is also foreseen. Development of 120 A radiation-hard power converters is also required.

6.5 Power converters and LHC machine availability

Global machine availability is affected by the pre-cycle needed to degauss the magnets and by the magnets ramp-down time. In the present LHC, the most limiting circuits are those of the inner triplet quadrupoles and of the main quadrupoles. All of these circuits are powered via one-quadrant converters, which are the cause for the long ramp-down time. Two upgrades can be envisaged if machine availability needs to be improved: i) replacement of these power converters with two-quadrant converter types; ii) use of external dump resistors to accelerate the discharge. As an illustration, by replacing the present 13 kA/18 V power converters of the main quadrupole circuits with two-quadrant 13 kA/±18 V power converters, the ramp down of the machine can be reduced by 30 minutes. As mentioned in Section 6.3, it is estimated that the replacement of both power converters powering the inner triplet quadrupole and main quadrupole circuits will increase the global availability of the machine by about 4%.